# Protocol architectures for IoT domains

Jelena Mišić, M. Zulfiker Ali, and Vojislav B. Mišić
Ryerson University
Toronto, ON, Canada M5B 2K3
Email: { jmisic, mzulfiker.ali, vmisic}@ryerson.ca

## Abstract

*In this work we discuss proxy architectures which interconnect IoT domains running CoAP with the rest of Internet including micro datacenters and other domains building scalable hierarchical architectures. We assume that CoAP domain is terminated by an IoT proxy with cache, and we investigate several design issues with respect to successful data transmission, round trip delay and energy consumption. We present performance data for the case when proxy autonomously maintains data freshness which clearly point to efficient design choices.*

## 1   INTRODUCTION

Efficient integration of IoT (Internet of Things) networks with Fog computing paradigm necessitates the deployment of IoT data concentrators to collect data from sensor nodes in IoT domains and deliver them to micro datacenters located at the Internet edge and close to clients. Those concentrators should have short access delay towards IoT nodes, preferably using single-hop communications over suitable short- to medium range wireless technology such as IEEE 802.15.4, Bluetooth LE, or IEEE 802.11ah. They provide data upon request from micro datacenter and, thus, act as proxies that store data in their cache and keep updating it with fresh data from servers (IoT nodes) that perform actual sensing. This approach relieves the micro datacenter from communicating directly with a multitude of IoT nodes and decreases its access time to sensed data. The concentrators (hereafter referred to as proxies) can even perform additional processing of cached data (e.g., classification or statistical processing) before sending the results to the micro datacenter, effectively converting the proxy into a kind of pico datacenter in the fog computing paradigm. Jointly, IoT nodes and proxy comprise the IoT domain such as smart city, smart grid, home monitoring and industrial plant infrastructures.

Another challenge in the design of IoT networks is the need to process large amounts of data efficiently and reliably, despite the limited computational and communication capabilities of IoT devices. To this end, the standard Internet protocol stack in IoT domains is replaced with a lightweight alternative stack that follows the Representational State Transfer (ReST) paradigm [1], which makes it upward compatible with the current Internet and ensures widespread acceptance. In the IoT protocol stack, human-readable HTML and XML messages are replaced by Efficient XML Interchange (EXI) which uses binary data [2]; HTTP running over TCP is replaced by much simpler Constrained Application Protocol (CoAP) running over UDP [3]; and IPv4/IPv6 is replaced by IPv6 and its adaptation for Low power Wireless Personal Area Networks (6LoWPAN) [4]. Conversion between those protocols, including IPv4-to-IPv6 address translation, is performed at the proxy. Fig. 1 shows the resulting architecture, with clients connected to the proxy either through the micro datacenter or directly.

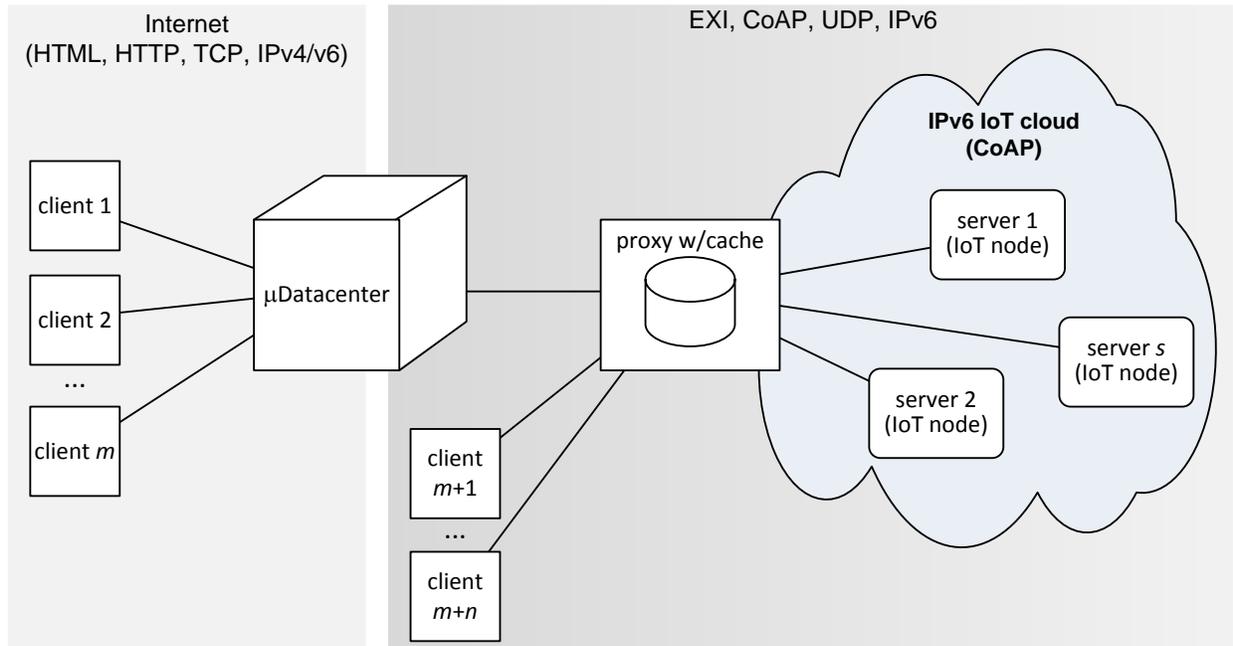

Figure 1. Topology of IoT domain at the network edge.

Performance of IoT proxies obviously depends on the communication technology as well as on other components of the network architecture. While Physical (PHY) and Medium Access Control (MAC) layers are typically constrained by the chosen technology and related hardware and energy considerations, IoT network designer is free to make design choices in areas such as cache maintenance strategy (reactive vs. proactive), unicast vs. multicast communications [3] [5], and use (or not) of observation-based mode in which the IoT node, after the initial poll from the proxy, continues to stream individual data readings [6]. These and other trade-offs involved in the choice of the communication paradigm are not well researched, with only a handful of studies available so far [7] [8].

In this paper we discuss three main design options for IoT domains running over CoAP: traditional POST/GET, multicast GET, and GET with observe option. The paper is organized as follows. Section 2 gives more details about CoAP and multicast CoAP in particular. Section 3 presents the model of multicasting proxy operation at the IoT proxy, while Section 3 presents three approaches towards building CoAP domain. Section 4 presents the performance comparison among domain design approaches, while Section 5 concludes the paper.

## 2. CONSTRAINED APPLICATION PROTOCOL CoAP AND MULTICAST COAP

### A. CONFIRMABLE AND NON-CONFIRMABLE TRANSFER

CoAP addresses the limitations of memory- and CPU-constrained nodes communicating using noisy and bandwidth-constrained wireless networks. It supports main methods supported in HTTP, namely GET, PUT, DELETE and POST. CoAP endpoints exchange requests and responses through messages exchanged asynchronously using UDP. As UDP does not provide reliable transfer, message reliability is supported through confirmable messages which require an acknowledgment or a reset message from the destination node. Reliability of confirmable messages is guaranteed by stop-and wait protocol in which re-transmission is triggered by lack of acknowledgment within the predefined timeout. A CoAP response will be carried either in an acknowledgment message or in a separate confirmable message. Duplicates are discarded based on message ID, while wrong parameters in the message are flagged with a reset message.

Non-confirmable messages will not be acknowledged although they may also receive a reset message upon error. Reliability can be supported through a data link layer protocol that supports acknowledged transfer. In the case of single-hop IEEE 802.15.4 network terminated with IoT proxy, non-confirmable transfer gives faster access to data compared to confirmable mode.

CoAP default delay parameters are given in Table 1.

Table 1. CoAP delay parameters

| parameter | default value |
|---|---|
| ACK timeout | 2s |
| ACK_RANDOM_FACTOR | 1.5 |
| MAX_RETRANSMIT | 4 |
| DEFAULT_LEISURE | 5s |
| MAX_TRANSMIT_SPAN | 45s |
| Processing Delay | 2s |
| MAX_RTT | 202s |
| freshness of data | $\overline{T_s} = 60 \text{ s}$ |

B. MULTICAST FEATURE

Multicast COAP (MCoAP) [3] [5] is an extension that improves scalability and efficiency in line with the lightweight concept of the CoAP protocol. IoT nodes which want to expose their resources can join one or more all-CoAP-node multicast addresses. All IoT nodes by default join the multicast group called 'All CoAP nodes' [9] with IPv4 address of `224.0.1.187`, and IPv6 address of `ff02::fd`, for the link-local scope, and `ff05::fd`, for the site local scope [10]. (Core link format described in [10] [11] also supports query filtering.) To support resource discovery in IoT domain, all nodes in the default multicast group expose their resources via default link '`/.well-known/core`'. They receive requests at the default UDP protocol port dedicated to CoAP.

Resource discovery is performed by sending a multicast GET (MGET) to the default link at default UDP protocol port [5] [12] [10] [11]. IoT nodes belonging to multicast group reply with links that are entry points to resource interfaces they host. Resource discovery may need to be performed periodically because nodes may join or leave the multicast group. MGET method is sent in a non-confirmable message to which IoT nodes reply in non-confirmable unicast with responses 2.05 (Content) or, if the resource is not found, with 4.04 (Not found); in the latter case, the IoT node may choose to withold the response [5].

MCoAP also brings challenges such as mitigating the congestion of replies and achieving reliability in receiving replies, including uniqueness of message IDs and tokens. Reliability can be established using acknowledged MAC layer, since MGET messages are nonconfirmable, while congestion mitigation requires that IoT nodes reply to the MGET after the so-called leisure time [3]. As leisure extends the response time for multicast and unicast GET methods, care must be taken that message IDs sent from the requesting node have a lifetime longer than the maximum multicast response time. Releasing the token value used to pair the query with the appropriate reply in multicast is another problem: namely, in unicast communication, the reception of reply from IoT node frees the token value, but in multicast it can be released only when most of the replies have been received. Recommended minimum token release value in [5] is 250 seconds, which puts a limit on the number of tokens to be maintained.

### C. OBSERVE FEATURE

Important addition to CoAP is the 'observe' feature [6]. In this scenario, a client ('observer') registers its interest to follow some physical variable at the IoT device ('subject'). If the subject accepts registration, it will notify the observer whenever the physical variable changes its value. The subject maintains the list of all its registered observers. One observer can register with multiple subjects in order to follow multiple physical variables. Data is flowing from the subject to the observer as a stream of 'notifications' which the subject has to send to all observers from the list.

Registration request is implemented as a GET request with observe option set and a unique token. Notification stream follows as the reply to registration GET with observe option and same token. Sequence numbers are provided in notifications so that re-order problem can be resolved. Notifications can be sent in confirmable or non-confirmable messages; the former have to be acknowledged by the client. Client can also de-register, i.e., cancel its interest in physical variable using GET request with the same token but with de-register bit set.

## 3. DOMAIN ARCHITECTURES

### A. PROXY CLASSIFICATIONS

IoT proxy collects and stores sensed data from IoT devices while maintaining the freshness of data stored in the cache. This can be achieved in a reactive or proactive mode, which is why the standard [3] classifies proxies into two broad classes, forward and reverse.

In the forward scheme, data acquisition is triggered by a GET request from the client which the proxy extends to the server (IoT node). Server responds with 2.05 (Content) or 4.04 (Not found), if the resource is not found; in the latter case response may not be sent due to congestion issues [12]. Therefore freshness of cached data depends only upon the dynamics of client queries.

In the reverse approach, servers update the cache with fresh data using POST or observe feature [6] In this case freshness of data depends on the data dynamics of IoT nodes, i.e., on parameterization of physical variables. Namely although raw sensing data may be periodical, it is more energy- and bandwidth-efficient if the data changes value upon some conditions. For example, a query such as `<coap://server/temperature/felt>` causes the physical variable to change state to 'cold', 'warm' and 'hot' whenever the sensed temperature reading is between given thresholds. However, parameterization of physical variable results in random times between cache updates. Time between updates can be estimated at the IoT node and sent with the data (in the Max_Age and ETag options) or it can be estimated at the proxy. In either case, randomness of data lifetime means that the proxy should actively maintain freshness of cached data. This leads to the so-called hybrid proxy scheme where user or proxy action is added to the normal updates from IoT nodes. Namely, if a user query finds that the record is obsolete, an additional GET request will be forwarded to IoT domain. Alternatively, the proxy may check data freshness itself and proactively refresh the data using unicast GET methods so that user queries find fresh data with high probability. Special type of hybrid proxy is the multicast-based one where data from IoT nodes are fetched using a MGET request. Cache can be refreshed using MGET reactively upon user request, or proactively by proxy upon expiration of freshness of one or more data records.

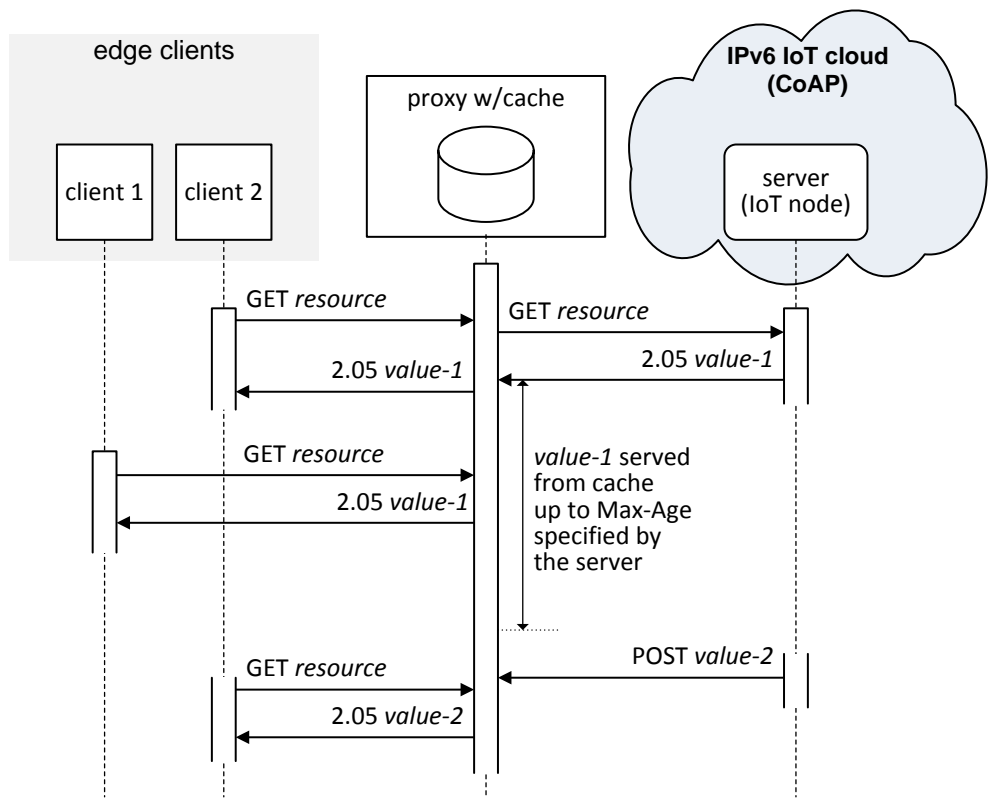

(a) Operation of POST/GET proxy.

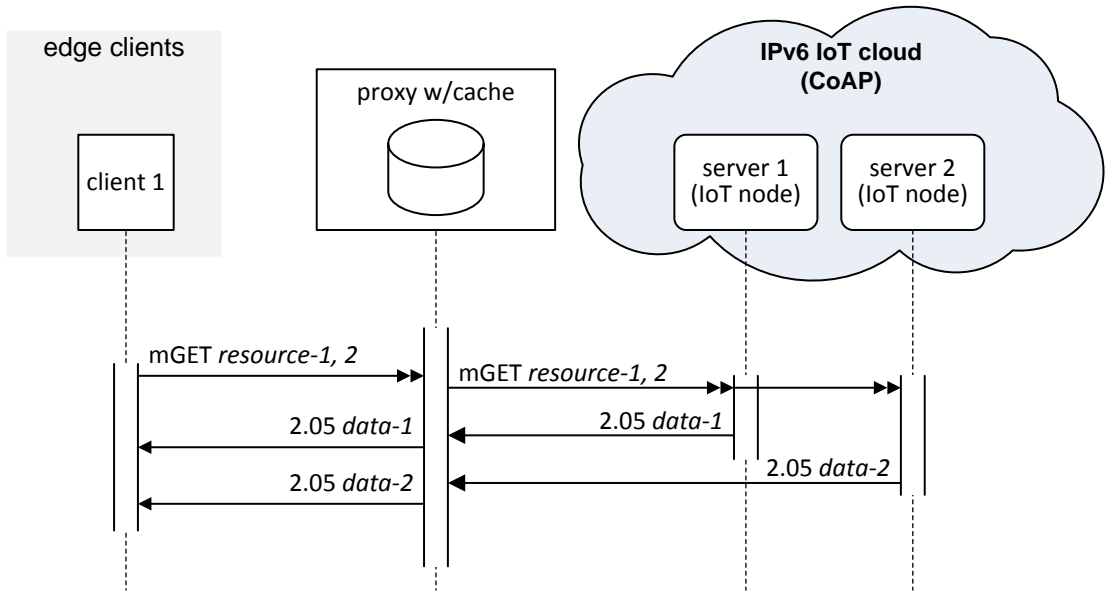

(b) Operation of multicast GET scheme.

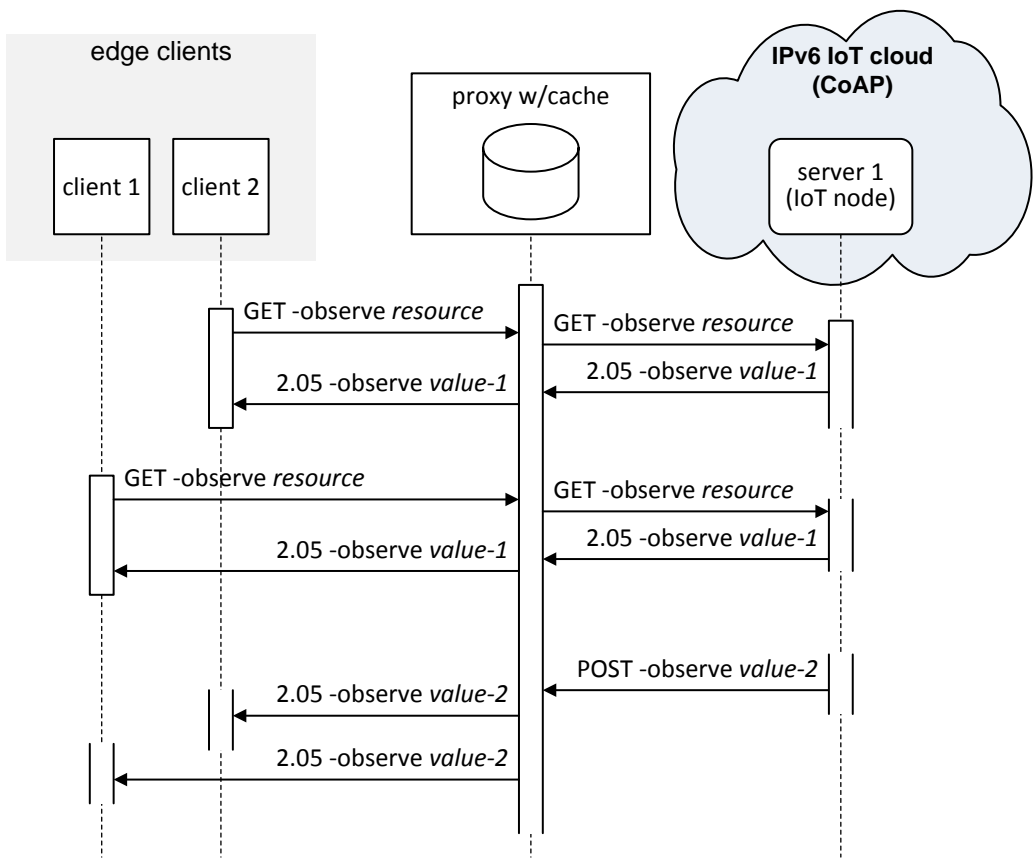

(c) Operation of GET -observe scheme.

Figure 2. Different CoAP techniques used for proxy operation.

Another classification of proxy operation may be based on CoAP techniques used.

In POST/GET based proxy, reactive operation uses a combination of POST methods sent from IoT nodes and occasional forwarded unicast GET from the user. Proactive operation is achieved when the proxy sends unicast GET requests towards IoT nodes in order to validate/refresh the data.

For multicast operated proxy, reactive operation consists of transmission of MGET requests towards group of nodes in IoT domain whenever a user unicast request finds record freshness below the threshold. Multicast maintenance of the cache is useful when data resources at the group of IoT nodes are parameterized in the same way, i.e., when data generation periods have similar distributions over nodes in IoT domain. Proactive proxy operation consists of regular checking of freshness of data records and transmitting multicast request towards IoT nodes when freshness of $k \in (1..n)$ nodes, where $n$ is the domain size, falls below limit. Efficiency of this design largely depends on the leisure period after which individual IoT node replies to MGET. This will be further discussed under congestion control.

For observe-based proxy, cache is regularly replenished using observing streams from IoT nodes. In proactive approach, the proxy checks data freshness and if necessary issues re-registration and validation GET request towards the IoT node with obsolete data. In reactive approach, a client request will cause re-registration GET request to be sent. Approach with observations puts lower load on IoT domain since a reply is not sent, unlike POST to which a reply to such as 2.01 (created) is necessary.

The operation of these schemes is schematically illustrated in Fig. 2.

## B. ESTIMATION OF THRESHOLD OF DATA FRESHNESS

CoAP documents [3] suggest that the IoT node can estimate the interval of data validity by sending values for Max_Age. Upon expiry of Max_Age value, proxy will further validate the record freshness by sending value of ETag option. This approach is suitable in the case when data generated by the IoT node have inter-generation time which is constant or varies between known minimal and maximal values. Under this scheme, proxy stores received Max_Age times and makes data validation requests when age of data record exceeds Max_Age.

It is more energy- and bandwidth-efficient for IoT nodes to deploy parameterization of sensed data [6] so that data is sent only upon certain conditions which results from execution of non-trivial queries. As a consequence such queries may result in random inter-generation time for data, which results in longer mean inter-generation time compared to transmission of raw (periodical) sensor data which will, in turn, reduce traffic intensity in IoT domain and energy consumption at IoT node.

However, random data inter-generation time makes estimation at the IoT node harder. In addition, it is inefficient to send it in Max_Age option to the proxy since total data lifetime will be affected by transmission time from the IoT node to proxy. Instead, it is better to estimate data lifetime at the proxy which is likely to have more computational resources anyway. Proxy will monitor and record time between arrivals of data for POST, multicast or observe-based methods. For high-end proxies, it may be possible to derive probability distribution of data interarrival time, while the lower-end ones may simply derive mean value and standard deviation using moving exponential averaging [13]. However, data inter-arrival time at proxy is not the same as data inter-generation time due to packet transmission time and, in case of multicast approach, leisure period. All these times are random and it is necessary to remove their impact (jitter) from inter-arrival time. As the since proxy cannot obtain information about transmission time from the IoT node, it must use information about round trip time ($RTT_p$) which can be obtained after sending GET request and receiving the reply. Depending on the design of the proxy, this can be done with a validation GET (if POST method is used), re-registration GET (for observe option), or MGET. In the case of multicast, the proxy should determine parameters of probability distribution of leisure time that should be deployed at IoT nodes, and convey this configuration value to the IoT domain in resource discovery multicast request. It can also be communicated whenever change of configuration of leisure time distribution is needed.

When probability distribution of pure inter-arrival time of data is found, proxy should determine data lifetime, i.e., Max_Age parameter. Simplest way to calculate it is as the sum of mean value and a certain number, say, $t$, of standard deviations. Choice of parameter $t$ is important since it balances real data freshness and traffic intensity in IoT domain. For example, large value of Max_Age will reduce the number of GET requests (of any kind) sent to IoT domain.

### C. CONGESTION CONTROL

Congestion control can be implemented at the proxy, IoT node, or both. A viable way of congestion control at the proxy is to dynamically change Max_Age parameter by adjusting the number of standard deviations in its calculation. This means that, under high traffic in IoT domain, Max_Age may increase which may temporarily affect freshness of data in the cache.

Congestion control at the IoT node prevents transmission of data regardless of transmission approach (POST, observe, reply to multicast) if the time from previous transmission is smaller than the round trip time $RTT_s$ measured at IoT node. This raises the question of how to measure round trip time at IoT node. In the case of POST based proxy, this is straightforward since every POST method will get a 2.01 (created) response from proxy.

For multicast and observe cases, this is more complex since replies to MGET and observation data do not receive any CoAP reply. In the case of multicast, the proxy should measure the number of received replies and if this number is too small, the period between multicast requests and leisure time for replies should be increased. In case of observe-based proxy, the solution has to be found in the balance between confirmable and non-confirmable CoAP transfer modes. Normally, the IoT node has a preference for non-confirmable transfer mode since the underlying MAC is reliable: both IEEE 802.15.4 and IEEE 802.11ah have acknowledged transfer with multiple re-transmissions in case of frame collision or noise error. However, to measure round trip time, some transmissions from IoT node have to be in confirmable mode. Portion of confirmable transmissions coming from IoT nodes has to be carefully estimated so that it does not much affect total traffic in the domain.

In any case, congestion control at the IoT node will affect estimation of data lifetime at the proxy, i.e., increase of drop rate at IoT node will increase Max_Age value and less validation GET requests will be sent. Therefore traffic in IoT domain will decrease and, consequently, congestion control at IoT node will drop fewer frames.

D. LEISURE TIME FOR MULTICAST BASED PROXY

Multicast-based proxy refreshes its cache using MGET requests. In this design, leisure time after which IoT node sends reply to MGET has significant impact on congestion. If this time is nonexistent or short all data replies from IoT nodes will be sent in short window. Although modern IoT technologies mostly deploy CSMA/CA at the MAC layer, contention windows are small and data packets could collide, leading to further re-transmissions. Therefore it is necessary to introduce additional random delay (leisure) at CoAP level after which the IoT node will submit the packet to MAC layer for transmission. Some of the techniques to create leisure time are as follows.

- Each node, at the network/MAC layer, could generate random individual leisure period at the boundaries of predefined slots, preferably matching backoff periods at the MAC layer. This time can have uniform or truncated geometric distribution since they can have simple implementations at IoT nodes. Probability distribution parameters have to be identical for all nodes unless some priority scheme among nodes is desirable. The period between MGET requests will be expressed in slots.
  Ratio between mean leisure period and the period between multicasts in this approach is constant and can be viewed as duty cycle which ranges between 0 and 1. Duty cycle is important parameter since, due to the randomness of leisure time and time between multicasts, it is possible that some nodes reply to current MGET request with $ID = n$ after the next request with $ID = n + 1$ is sent by the proxy. This leads to the problem of determining the token release time after which late replies to a MGET with token $x$ will not be recognized. Obviously, large duty cycle will alleviate congestion but it will also require large token release times.
  Due to randomness of time between MGET requests it is better to express token release time as the number of MGET requests following the target one, after which token can be released. This number can be derived by specifying the probability that token cannot be matched.
- Previous scheme can be made adaptive by allowing duty cycle to be proportional to the number of nodes or traffic intensity in the IoT domain. This scheme is suitable for nodes which follow some sleeping scheme. It is also necessary for implementation of congestion control at the proxy, but the proxy must measure the number of responses per request in order to estimate the number of active IoT nodes and to transmit parameters of leisure time in dedicated multicast transmissions.
- If IoT nodes use CSMA/CA MAC layer, it is also possible to integrate the leisure period in the backoff process at the IoT node, i.e., by increasing the backoff window in proportion to the number of nodes. Indeed backoff windows for popular IEEE 802.15.4 technology start with values smaller than 16, and cannot accommodate simultaneous transmissions from large numbers of IoT nodes. However, if the receiving antenna is on during the backoff countdown (as is the case in IEEE 802.15.4), longer backoff windows lead to larger energy consumption at IoT node radio subsystem.

E. LAYERED DESIGN OF PROXY

Due to device limitations, CoAP was designed for IoT devices with a single UDP socket. Currently, only a single thread is allocated to deal with UDP socket and a small number of threads deal with CoAP methods such as GET, PUT, POST and DELETE [14]. This may impose performance problems at CoAP proxies where multi-threading and real time scheduling becomes necessary. Proxy has Internet and IoT sides, each running on separate core. Each proxy side has layered architecture consisting of caching application, CoAP, security (DTLS), transport (UDP), and network/MAC layer. Layers are served by threads and interfaced with message queues. Addition of multiple queues per layer allows service differentiation either per user or per physical variable. Depending on the proxy design additional differentiation can be achieved, for example by using different duty cycle for multicast proxy. Number of threads per layer depends on operating system and computational resources of the proxy. Initial implementations of layered/staged architecture are presented in [15] [14].

## 4. PERFORMANCE COMPARISON

The POST/GET, MGET and observe/GET schemes have disjoint sets of design parameters which makes direct comparisons hard. To allow for fair comparison, we have considered proactive proxy design where proxy maintains freshness of the cache and user queries are not forwarded to the IoT domain. Number of protocol layers at the proxy was set to three, with each layer having exponentially distributed thread execution time with mean of 5ms.

Lifetime of physical variables was exponentially distributed with mean of 60s and limit for data freshness was set to 60s. CoAP method transmissions were non-confirmable and congestion control was not implemented since it would introduce performance heterogeneity. The common point in three approaches is that probability of outdated record is $e^{-1}$ for single physical variable. For multicasting proxy leisure duty cycle was proportional to the IoT cluster size.

We have investigated performance against IoT domain size in the range between 50 and 500 active nodes. Nodes were equipped with IEEE 802.15.4 communication interfaces with packet size 127B. Energy consumption of the communication interface per backoff period was set to $\omega_s = 18.2$nJ, $\omega_r = 17.9\mu$J and $\omega_t = 15.8\mu$J, during inactive, receiving and transmitting (at 0dBm) state, respectively.

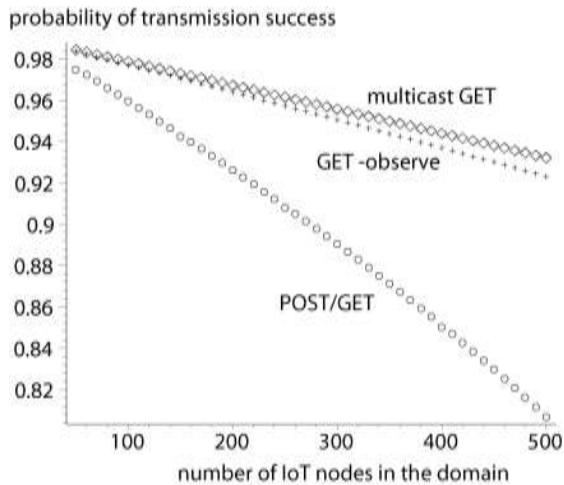

(a) Probability of transmission success.

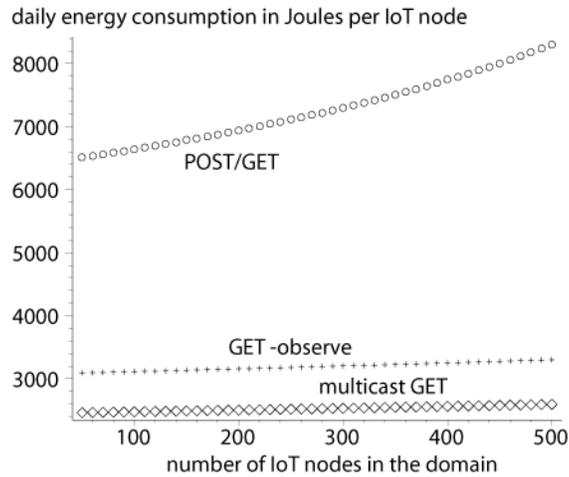

(b) Daily energy consumption per IoT node in Joules.

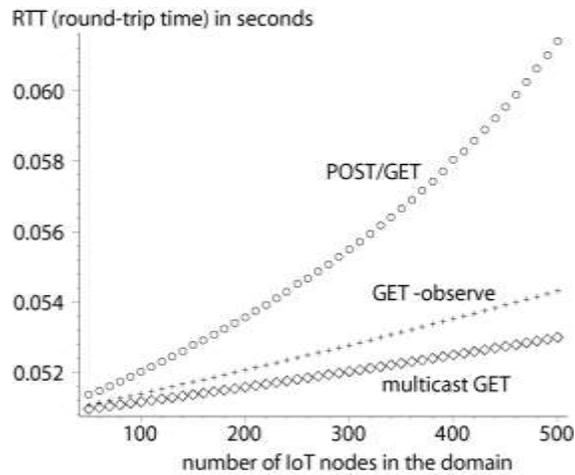

(c) Round-trip time in the Iot in domain in seconds.

Figure 3. Main performance indicators for POST/GET (circle), multicast GET (diamond), and GET -observe (cross) schemes.

Probability of single transmission success, energy consumption at communication interface of the IoT node, and round-trip time in the cluster are shown in Fig. 3. Curves are denoted with circle, diamond and cross symbols for POST/GET, MGET and observe/GET proxy designs respectively.

While results for POST/GET and observe/GET proxy have stationary behavior in time, results for MGET proxy exhibit peak and silent regimes. Therefore we have presented only mean values of performance descriptors for multicast proxy, noting that peak values occur soon after MGET request is received and depend on the chosen duty cycle.

Round-trip time is important since its value affects overall data freshness. Namely when proxy detects outdated record(s) and sends validation GET/MGET requests, data will be outdated until the reply arrives. Round trip time incorporates uplink and downlink processing times through proxy layers as well as uplink and downlink waiting and transmission times at communication interfaces (proxy and IoT node).

Performance descriptors show that POST/GET approach has lowest probability of success, highest

RTT and highest daily energy consumption. Multicast and observe-based proxies have similar performance with the former having a slight advantage. Indeed POST/GET scheme can accommodate at most 300 IoT nodes while other two designs can accommodate between 500-600 nodes. Reactive proxy designs have similar behavior since total performance depends on the traffic intensity in the IoT domain which is lightest for multicast design and highest for POST/GET design.

## 5. CONCLUSION

In this paper we have discussed possible design approaches for IoT domains that run CoAP and interface to the Internet through a proxy containing data cache. This design enables scalable fog computing paradigm since domains can further be interconnected in hierachical manner allowing micro datacenters to access the data quickly. We have investigated three communication paradigms in the IoT domain namely POST/GET, multicast GET and observation/GET based, under similar design options. Results show that multicast based proxy exhibits the best performance, followed by observe/GET and POST/GET designs.